\documentclass{llncs}
\usepackage{amsmath}
\usepackage{amssymb}
\usepackage{subfigure}
\usepackage{epic}
\newcommand{\Nb}{{\mathbb{N}}}

\newcommand{\B}{{\cal B}}

\newcommand{\R}{{\cal R}}

\newcommand{\N}{{\cal N}}

\begin{document}

\title{Algorithmic analogies to Kamae-Weiss theorem on  normal numbers}
\author{Hayato Takahashi
\institute{The Institute of Statistical Mathematics, \\10-3 Midori-cho, Tachikawa, Tokyo 190-8562, Japan. \\ Email: hayato.takahashi@ieee.org}}

\maketitle

\begin{abstract}
In this paper we study subsequences of random numbers.
In Kamae (1973),  selection functions that depend only on coordinates are studied, and their necessary and sufficient condition for the selected sequences to be normal numbers  is given.
In van Lambalgen (1987), an algorithmic analogy to the theorem is conjectured in terms of algorithmic randomness and Kolmogorov complexity.
In this paper, we show different algorithmic analogies to the theorem.
\end{abstract}

\section{Introduction}\label{sec-intro}
In this paper we study subsequences of random numbers.
A function from sequences to their subsequences is called selection function. 
 In Kamae \cite{kamae73}  selection functions that depend only on coordinates are studied, and their necessary and sufficient condition for the selected sequences to be normal numbers is given.
In the following we call the theorem Kamae-Weiss (KW) theorem on normal numbers since a part of the theorem is shown in  Weiss \cite{weiss71}.
In van Lambalgen \cite{lambalgen87}, an algorithmic analogy to KW theorem is conjectured in terms of algorithmic randomness and complexity \cite{chaitin75,Kol65,martin-lof66,solomonoff64}. 
In this paper we show two algorithmic analogies to KW theorem.

Let \(\Omega\) be the set of infinite binary sequences.
For \(x,y\in\Omega\), let \(x=x_1x_2\cdots\) and \(y=y_1y_2\cdots,\ \forall i\ x_i, y_i\in\{0,1\}\).
Let \(\tau:\Nb\to\Nb\) be a strictly increasing function such that 
\(\{ i\mid y_i=1\}=\{\tau(1)<\tau(2)<\cdots\}\).
If \(\sum_i y_i=n\) then \(\tau(j)\) is defined for \(1\leq j\leq n\).
For \(x,y\in\Omega\) let \(x/y\) be the subsequence of \(x\) selected at \(y_i=1\), i.e.,
\(x/y=x_{\tau(1)}x_{\tau(2)}\cdots\).
For example, if \(x=0011\cdots,\ y=0101\cdots\) then \(\tau(1)=2, \tau(2)=4\) and \(x/y=01\cdots\).
For finite binary strings  \(x_1^n:=x_1\cdots x_n\) and \(y_1^n:=y_1\cdots y_n\),
\(x_1^n/y_1^n\) is defined similarly. 
Let \(S\) be the set of finite binary strings and \(|s|\) be the length of \(s\in S\).
For \(s\in S\) let \(\Delta(s):=\{s\omega | \omega\in\Omega\}\), where \(s\omega\) is the concatenation of \(s\) and \(\omega\).
Let \((\Omega,\B, P)\) be a probability space, where \(\B\) is the sigma-algebra generated by \(\Delta(s), s\in S\).
We write \(P(s):=P(\Delta(s))\).

In Kamae \cite{kamae73}, it is shown that the following two statements are equivalent under the assumption that \(\liminf \frac{1}{n}\sum_{i=1}^n y_i>0\):
\begin{theorem}[KW]\hfill\\
(i) \(h(y)=0\).\\
(ii) \(\forall x\in \N\ x/y\in \N\),\\
where \(h(y)\) is  Kamae entropy \cite{brudno83,lambalgen87} and \(\N\) is the set of binary normal numbers.
\end{theorem}
A probability \(p\) on \(\Omega\) is called cluster point if there is a sequence \(\{n_i\}\)
\[\forall s\in S\ p(s)=\lim_{i\to \infty} \#\{ 1\leq j\leq n_i \mid x_j\cdots x_{j+|s|-1}=s\}/n_i.\]
From the definition,  the cluster points are stationary measures. 
Let \(V(x)\) be the set of cluster points defined from \(x\).
 From a standard argument we see that \(V(x)\ne\emptyset\) for all \(x\).
Kamae entropy is defined by
\[h(x)=\sup\{ h(p)\mid p\in V(x)\},\]
where \(h(p)\) is the measure theoretic entropy of \(p\).
If \(h(x)=0\), it is called completely deterministic, see \cite{kamae73,weiss71,weiss00}.
The part (i)\(\Rightarrow\) (ii) is appeared in \cite{weiss71}.

As a natural analogy,  the following equivalence (algorithmic randomness version of Kamae's theorem) under a suitable restriction on \(y\) is conjectured   in van Lambalgen \cite{lambalgen87},\\
(i) \(\lim_{n\to\infty}K(y_1^n)/n=0.\)\\
(ii) \(\forall x\in\R\ x/y\in\R\),\\
where \(K\) is the prefix Kolmogorov complexity and \(\R\) is the set of Martin-L\"of random sequences with respect to the uniform measure (fair coin flipping), see \cite{LV2008}.
Note that  \(\lim_{n\to\infty}K(y_1^n)/n=h, P-a.s.,\) for ergodic \(P\) and its entropy \(h\), see \cite{brudno83}.
\section{Results}
In this paper, we show two algorithmic analogies to KW theorem.
The first one is a Martin-L\"of randomness analogy and the second one is a complexity rate analogy to KW theorem, respectively. 
In the following, 
\(P\) on \(\Omega\) is called computable if there is a computable function \(A\) such that \(\forall s, k\ |P(s)-A(s,k)|<1/k\).
For \(A\subset S\), let \(\tilde{A}:=\cup_{s\in A}\Delta(s)\).
A recursively enumerable (r.e.) set \(U\subset \Nb\times S\) is called (Martin-L\"of) test with respect to \(P\) if 1) \(U\) is r.e., 2) \(\tilde{U}_{n+1}\subset \tilde{U}_n\) for all \(n\), where \(U_n=\{s : (n,s)\in U\}\), and
3) \(P(\tilde{U}_n)<2^{-n}\). 
A test \(U\) is called universal if for any other test \(V\), there is a constant \(c\) such that \(\forall n\ \tilde{V}_{n+c}\subset \tilde{U}_n\).
In \cite{martin-lof66}, it is shown that a universal test \(U\) exists if \(P\) is computable and the set \((\cap_{n=1}^\infty \tilde{U}_n)^c\) is called the set of Martin-L\"of random sequences with respect to \(P\).

Our first algorithmic analogy to the KW theorem is the following. 
\begin{proposition}
Suppose that \(y\) is Martin-L\"of random with respect to some computable probability \(P\) and \(\sum_{i=1}^\infty y_i=\infty\).
Then the following two statements are equivalent:\\
(i) \(y\) is computable.\\
(ii) \(\forall x\in\R\ x/y\in\R^y\),\\
where \(\R^y\) is the set of Martin-L\"of random sequences with respect to the uniform measure relative to \(y\). 
\end{proposition}
Proof) (i)\(\Rightarrow\) (ii).
Since \(\sum_{i=1}^\infty y_i=\infty\) we have   \(\forall s\ \lambda \{ x\in\Omega\mid s\sqsubset x/y\} =2^{-|s|}\), where \(\lambda\) is the uniform measure.
Let \(U\) be a universal test with respect to \(\lambda\) and \(y(s)\subset S\) be a finite set such that \(\{ x\in\Omega\mid s\sqsubset x/y\}= \tilde{y}(s)\).
Then \(y(s)\) is computable from \(y\) and \(s\), and hence \(U^y:=\{ (n, a)\mid a\in y(s), s\in U_n\}\) is a test if \(y\) is  computable.
We have \(x\in \tilde{U}_n^y\leftrightarrow x/y\in \tilde{U}_n\).
(Intuitively  \(U^y\) is a universal test on subsequences selected by \(y\)).
Then 
\[
x\in\R  \to x\notin \cap_n\tilde{U}_n^y \leftrightarrow x/y \notin \cap_n\tilde{U}_n \leftrightarrow x/y\in\R.
\]
Since \(y\) is computable, \(\R^y=\R\) and we have (ii).

Conversely,  suppose  that \(y\) is a  Martin-L\"of random sequence with respect to a computable \(P\) and is not computable. 
From Levin-Schnorr theorem, we have
\begin{equation}\label{ls}
\forall n\ Km(y_1^n)=-\log P(y_1^n)+O(1),
\end{equation}
where \(Km\) is the monotone complexity. 
Throughout the paper, the base of logarithm is 2.
By applying arithmetic coding to \(P\), there is a sequence \(z\) such that \(z\) is computable from \(y\) and
\(y^n_1\sqsubset u(z^{l_n}_1),\ l_n= -\log P(y^n_1)+O(1)\) for all \(n\), where \(u\) is a monotone function and we write \(s\sqsubset s'\) if \(s\) is a prefix of \(s'\).
Since \(y\) is not computable, we have \(\lim_n l_n=\infty\). 
From (\ref{ls}), we see that \(\forall n\ Km(z_1^{l_n})=l_n+O(1)\).
We show that if \(y\in\R\) then \(\sup_n  l_{n+1}-l_n  <\infty\).
Observe that if \(y\in\R\) then \(\forall n\ P(y_1^n)>0\) and 
\begin{align*}
\sup_n   l_{n+1}-l_n  <\infty\leftrightarrow \sup_n -\log P(y_{n+1}\mid y_1^n)<\infty & \leftrightarrow \inf_n P(y_{n+1}\mid y^n_1)>0\\
& \leftrightarrow \liminf_n P(y_{n+1}\mid y^n_1)>0.
\end{align*}
Let  \(U_n:=\{ s \mid P(s \mid s_1^{|s|-1})<2^{-n}\}\). Then \(P(\tilde{U}_n)<2^{-n}\) and  \(U:=\{ (n, s)\mid s\in U_n\}\) is a r.e.~set. 
Since \(y\in\limsup_n \tilde{U}_n\leftrightarrow \liminf_n P(y_{n+1}\mid y_1^n)=0\), if \(y\in\R\) then \(\sup_n   l_{n+1}-l_n  <\infty\).
(If \(U\) is r.e.~and \(P(\tilde{U}_n)<2^{-n}\) then \(\R\subset (\limsup_n \tilde{U}_n)^c\), see \cite{shen89}.)
Since  \(\forall n\ Km(z_1^{l_n})=l_n+O(1)\) and \(\sup_n   l_{n+1}-l_n  <\infty\), we have \(\forall n\ Km(z_1^n)=n+O(1)\) and \(z\in\R\).
Since \(z\) is computable from \(y\) we have \(z/y\notin \R^y\).
\qed

Note that if \(y\) is computable then \(y\) is a Martin-L\"of random sequence with respect to a computable measure that has positive probability at \(y\).

In order to show the second analogy, we introduce another notion of randomness. 
We say that \(y\) has maximal complexity rate with respect to \(P\) if 
\begin{equation}\label{weak}
\lim_{n\to\infty}\frac{1}{n} K(y_1^n)=\lim_{n\to\infty}-\frac{1}{n}\log P(y_1^n),
\end{equation}
i.e., both sides exist and are equal. 
For example, \(y\) has maximal complexity rate with respect to the uniform measure (i.e., \(P(s)=2^{-|s|}\) for all \(s\)) if\\
  \(\lim_{n\to\infty} K(y_1^n)/n=1\).
If \(y\) is Martin-L\"of random sequences with respect to a computable ergodic \(P\)  then
from upcrossing inequality for the Shannon-McMillan-Breiman theorem \cite{hochman2009}, the right-hand-side of (\ref{weak}) exists (see also \cite{vyugin98}) and from
(\ref{ls}), we see that (\ref{weak}) holds i.e., \(y\) has maximal complexity rate w.r.t. \(P\). 

\begin{proposition}\label{main}
Suppose that \(y\) has maximal complexity rate with respect to a computable probability and \(\lim_n \frac{1}{n}\sum_{i=1}^n y_i>0\).
Then the following two statements are equivalent:\\
(i) \(\lim_{n\to\infty}\frac{1}{n}K(y_1^n)=0.\)\\
(ii) \(\forall x\ \lim_{n\to\infty}\frac{1}{n}K(x_1^n)=1\to \lim_{n\to\infty} \frac{1}{|x_1^n/y_1^n|}K(x_1^n/y_1^n |y_1^n )=1.\)
\end{proposition}
Proof)\\
(i) \(\Rightarrow\) (ii)\\
Let \(\bar{y}:=\bar{y}_1\bar{y}_2\cdots\in\Omega\) such that \(\bar{y}_i=1\) if \(y_i=0\) and \(\bar{y}_i=0\) else for all \(i\).
Since 
\[ | K(x_1^n)-K(x_1^n | y_1^n) | \leq K(y_1^n)+O(1)\]
 and 
 \begin{align*}
& K(x_1^n | y_1^n)=K(x_1^n/ y_1^n, x_1^n/ \bar{y}_1^n | y_1^n) +O(1),
\end{align*}
if \(\lim_{n\to\infty}K(y_1^n)/n=0\) and \(0<\lim_n \frac{1}{n}\sum_{i=1}^n y_i<1\) then 
we have
\begin{align*}
& \lim_{n\to\infty}K(x_1^n)/n=1\\
& \Rightarrow \lim_{n\to\infty} \frac{1}{n} K(x_1^n/ y_1^n, x_1^n/ \bar{y}_1^n | y_1^n)=1\\
& \Rightarrow \lim_{n\to\infty} \frac{1}{n}(K(x_1^n/ y_1^n| y_1^n)+K(x_1^n/ \bar{y}_1^n | y_1^n))=1\\
& \Rightarrow \lim_{n\to\infty} \frac{n_1}{n}\frac{1}{n_1}K(x_1^n/ y_1^n| y_1^n)+\frac{n-n_1}{n}\frac{1}{n-n_1}K(x_1^n/ \bar{y}_1^n | y_1^n)=1\\
& \Rightarrow \lim_{n\to\infty} \frac{1}{n_1}K(x_1^n/y_1^n |y_1^n )=1\text{ and }\lim_{n\to\infty}\frac{1}{n-n_1}K(x_1^n/ \bar{y}_1^n | y_1^n)=1.
\end{align*}
where \(n_1=|x_1^n/y_1^n|=\sum_{i=1}^n y_i\).
Similarly, if \(\lim_{n\to\infty}K(y_1^n)/n=0\)  and \\
\(\lim_n \frac{1}{n}\sum_{i=1}^n y_i=1\) then we have 
\( \lim_{n\to\infty} \frac{1}{n_1}K(x_1^n/y_1^n |y_1^n )=1\).\\
(ii) \(\Rightarrow\) (i)\\
Suppose that 
\begin{equation}\label{wls}
\lim_{n\to\infty}\frac{1}{n} K(y_1^n)=\lim_{n\to\infty}-\frac{1}{n}\log P(y_1^n)>0,
\end{equation}
for a computable \(P\).
Let \(l_n\) be the least integer greater than \(-\log P(y_1^n)\).
Then by considering arithmetic coding, there is \(z=z_1z_2\cdots\in\Omega\) and a monotone function \(u\)  such that 
\(y^n_1\sqsubset u(z^{l_n}_1)\).  By considering optimal code for \(z^{l_n}_1\) we have \(Km(y^n_1)\leq Km(z^{l_n}_1)+O(1)\).
From  (\ref{wls}), we have \(\lim_n Km(y^n_1)/l_n=\lim_n  Km(z^{l_n}_1)/l_n=1\).
For \(l_n\leq t \leq l_{n+1}\), we have \(Km(z^{l_n}_1)/l_{n+1}\leq Km(z^t_1)/t\leq Km(z^{l_{n+1}}_1)/l_n\).
From (\ref{wls}), we have \(\lim_n l_{n+1}/l_n=1\), and hence \(\lim_n Km(z^n_1)/n=\lim_n K(z^n_1)/n=1\).

Since 1) \(z^{l_n}_1\) is computable from \(y^n_1\), 2)  \(\lim_n l_n/n>0\) by (\ref{wls}), and \\
3)  \(\lim_n \frac{1}{n}\sum_{i=1}^n y_i>0\),  we have \(\limsup_{n\to\infty} \frac{1}{|z_1^n/y_1^n|}K(z_1^n/y_1^n | y_1^n)<1\).
\qed

\begin{example}
Champernowne sequence satisfies the condition of the proposition and (i) holds, however its Kamae-entropy is not zero. 
\end{example}

\begin{example}
If \(y\) is a Sturmian sequence  generated by an irrational rotation model with a computable parameter \cite{kamaeTakahashi,takahashiAndaihara} then \(y\) satisfies the condition of the proposition and (i) holds. 
\end{example}

\section{Discussion}
Both proofs of Proposition 1 and 2 have similar structure, i.e., the part (i) \(\to\) (ii) are straightforward
 and in order to show the converse, 
we construct random sequences (in the sense of Proposition 1 and 2, respectively) by compression. 

We may say that Proposition 1 is a Martin-L\"of randomness analogy and Proposition 2 is a complexity rate analogy to KW theorem, respectively. 
These results neither prove nor disprove the conjecture of van Lambalgen.
However Martin-L\"of randomness and complexity rate randomness give different  classes of randomness, and 
a curious point of the conjecture is that it states equivalence of statements described in terms of  them. 

As stated above we proved our propositions by constructing random sequences.
In \cite{dowe2011} pp.962,  a different direction is studied, i.e.,
a sequence that is not predicted by MML with respect to finite order Markov processes is considered. 
Such a sequence is called {\it red herring sequence} \cite{dowe2011} and considered to be a non-random sequence with respect to MML and finite order Markov processes, in the sense that 
MML cannot find a finite order Markov model for that sequence.

\begin{center}
{\bf Acknowledgement}
\end{center}
The author thanks Prof.~Teturo Kamae (Matsuyama Univ.) for discussions and comments.

\end{document}